\documentclass[doublecol]{epl2}
\usepackage[T1]{fontenc}
\pdfoutput=1
\usepackage[latin1]{inputenc}
\usepackage{geometry}
\usepackage[british]{babel}
\usepackage{amsmath}
\usepackage{amssymb}
\usepackage{graphicx}
\usepackage{esint}
\usepackage[normalem]{ulem}
\usepackage[usenames]{xcolor}
\newcommand{\red}{\color{black}}
\newcommand{\blue}{\color{black}}
\newcommand{\magenta}{\color{black}}

\title{	On the genealogy of branching random walks and of directed polymers}

\author{Bernard Derrida\inst{1,2} \and Peter Mottishaw\inst{3}}
\shortauthor{B. Derrida and P. Mottishaw}

\institute{                    
	\inst{1} Coll\`{e}ge de France,	11 Place Marcelin Berthelot, 75005 Paris, France\\
	\inst{2} LPS,  \'{E}cole Normale Sup\'{e}rieure, 24 rue Lhomond, 75005 Paris, France\\
	\inst{3} School of Physics and Astronomy, University of Edinburgh, James Clerk Maxwell Building, Peter Guthrie Tait Road, Edinburgh EH9 3FD, UK
}

\abstract{
	 It is well known that the mean field theory of directed polymers in a random medium  exhibits  replica symmetry breaking with a distribution of overlaps which consists of two delta functions. Here we show that the leading finite size correction to this distribution of overlaps has a universal character which can be  computed explicitly. Our results can also be interpreted as genealogical properties of branching Brownian motion or of branching random walks.
}
\begin{document}

\maketitle

\section{ Introduction}

{\blue
The study of branching Brownian motion and of  branching random walks is central in the theory of probability \cite{mckean,Bra,lal,shi} and 
    {\red appears in   several  physical contexts \cite{kra1,maj1,maj2,munier,ramola,montanari}}. } Here we focus on {\red its} revelance    in the  mean  field theory of directed polymers in a random medium \cite{HZ,spohn,hu,Ar1}.
 This mean field version 
is an example of a disordered system which exhibits {\magenta replica symmetry breaking} in its low temperature phase, {\red \cite{MP0,MP,parisi}}. {\magenta In the models considered here only one step of replica symmetry breaking is required in the thermodynamic limit} \cite{MPV}.
Our  motivation {\magenta here} is to determine  how this broken symmetry is affected by finite size  fluctuations.
{\blue In the present work we will show  } that the one step {\magenta replica symmetry breaking} is smoothed in a universal way  for which  one can obtain an explicit analytic expression.

\begin{figure}
\onefigure[width=7cm]{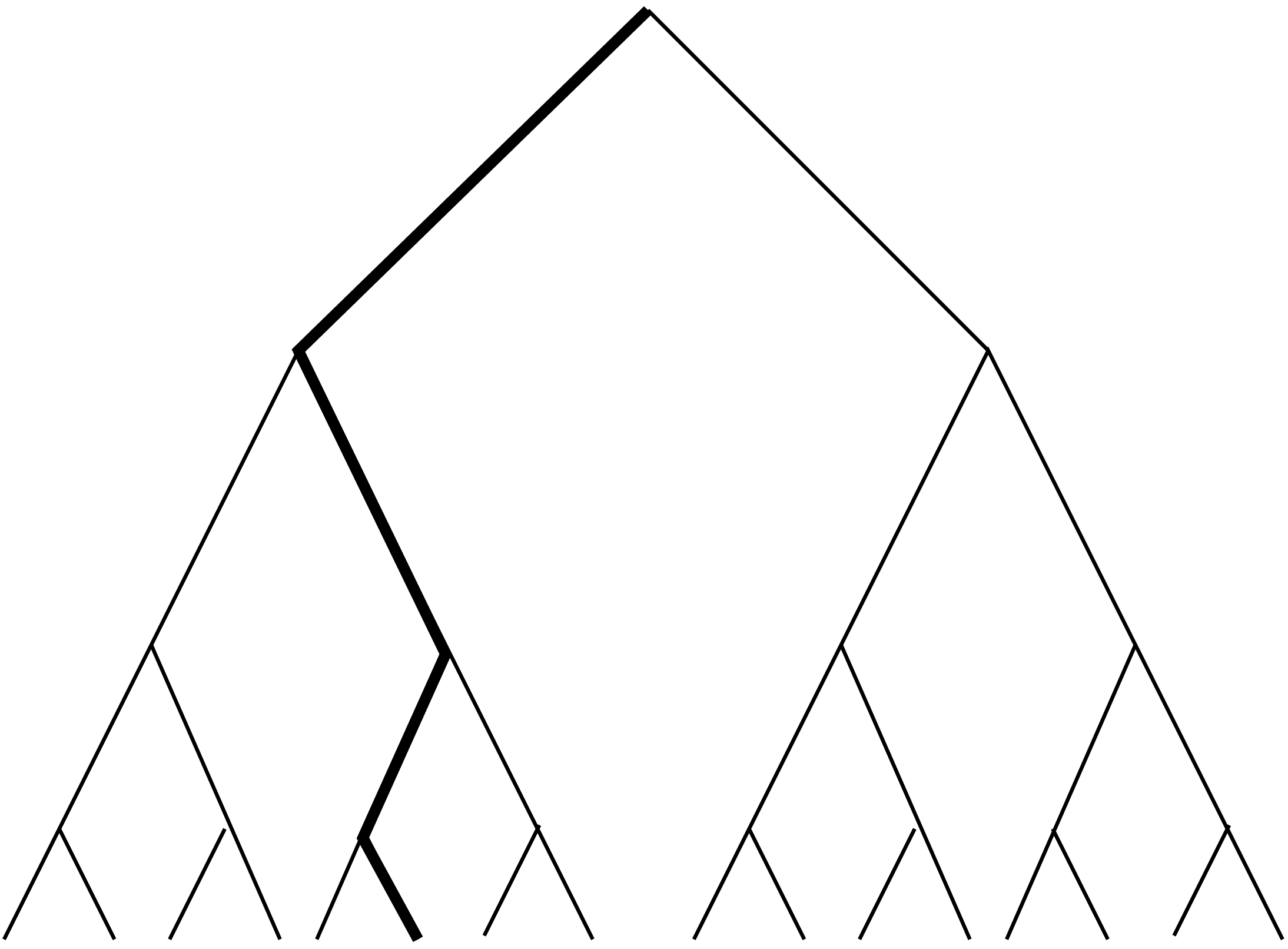}
\caption{ A binary tree of height $t$. On each edge there is a random energy $\epsilon_b$.
The configurations of a directed polymer are the $2^t$ paths connecting the top of the tree to its bottom.}
\label{tree-fig}
\end{figure}

To study the problem of directed polymers in its mean field {\red version},  one  may consider a binary tree of height $t$ as on figure \ref{tree-fig}. On each edge $b$ of this tree  there is a random energy $\epsilon_b$ distributed according to a given distribution $\rho(\epsilon)$. Then the possible configurations of the polymer are the $2^t$ paths of length $t$ connecting the top of the tree to its bottom. To each of these paths $i$ one associates an energy $X_i(t)$ which is simply the sum of the energies of all the bonds visited by this path
$$ X_i(t) = \sum_{b \in i} \epsilon_b \ . $$
The partition function is then given by
$$Z_t = \sum_{i=1}^{\blue {\cal N}(t) } e^{-\beta X_i(t)}$$
where the sum is over the {\blue${\cal N}(t)=2^t$} configurations of the directed polymer, {\blue  $\beta$ is the } inverse temperature and 
the weight $W_i$  of a path is 
\begin{equation}
 W_i= \frac{e^{-\beta X_i(t)}}{Z_{t}} \ . 
\label{Wi}
\end{equation}

For  a general distribution $\rho(\epsilon)$  {\blue of the energies $\epsilon_b$} it is known \cite{rio} that there is a phase transition  at $\beta_c$
given by the value of $ \beta$ which minimizes  the function $v(\beta)$ defined by
\begin{equation}
\label{vbeta}
 v(\beta) = 
\frac{g(\beta) + \log 2}{\beta}
\end{equation}
where
\begin{equation}
g(\beta) = \log \left[ \int \rho(\epsilon) e^{-\beta \epsilon} d \epsilon  \right]  
\label{gbeta}
\end{equation}

i.e. {\blue $\beta_c$ is solution of }
{\red 
\begin{equation}
v'(\beta_c)=0 \ .
\label{vprime}
\end{equation}
}
 It is also known that in
 the low temperature phase {\blue ($\beta> \beta_c$)}, the extensive part of the entropy vanishes and that the partition function
is dominated by the configurations whose difference of energy with the ground state is non extensive.

In the directed polymer problem on a tree the overlap $q_{i,j}$ between two paths is simply the fraction of their length {\magenta that the two paths share. So that,} $q_{i,i}=1$, while
$$q_{i,j}=\frac{r}{t}$$
if the two paths $i$ and $j$ have $r$ bonds in common. {\red 
The  distribution of overlaps $P(q)$ is then} defined by
$$
P(q) = \left\langle  \sum_{i,j} W_i \ W_j \  \ \delta(q_{i,j} -q) \right\rangle
$$
where $\langle . \rangle$ denotes an average over {\red all} the random energies $\epsilon_b$.

It is well established \cite{spohn,rio,Ar1} that in the low temperature phase ($ \beta > \beta_c)$
and in the limit $t \to \infty$,  this distribution $P(q)$ is {\red the}  sum of two $\delta$ functions
\begin{equation}
P(q) = \frac{\beta_c}{\beta} \ \delta(q) \ + \ \left( 1 - \frac{\beta_c}{\beta} \right)\  \delta(q-1)  \ . 
\label{P(q)}
\end{equation}
{\blue 
{\magenta In the low temperature phase} the distribution $P(q)$ of overlaps is non-trivial (in the sense {\red that, in the large $t$ limit,} it does not reduce to a single $\delta$  function). {\magenta This} is one of the signatures of {\magenta replica symmetry breaking} \cite{MPV}. The mean field directed polymer problem is one of the simplest examples of such a broken symmetry: {\red it exhibits} a one step broken replica symmetry} {\red since the  distribution of overlaps $P(q)$  consists  of a sum of  two $\delta$ functions (\ref{P(q)}).  
In the present work we study the leading finite size corrections to (\ref{P(q)}).  We will} show that for large {\red $t$,
 $0<q<1$}  and $\beta>\beta_c$
\begin{equation}
P(q) \simeq \frac{1}{\sqrt{t}} \frac{{\magenta 1}}{\beta}
\sqrt{ \frac{1}{2 \pi  \beta_c v''(\beta_c)}}
\ \big( q (1-q) \big)^{-\frac{3}{2} }
\label{r1}
\end{equation}

and similarly when $\beta  = \beta_c$

\begin{equation}
P(q) \simeq \frac{1}{\sqrt{t}} {{\magenta \frac{1}{\beta_c}}}
\sqrt{
	\frac{ 1}{2 \pi \beta_c   v''(\beta_c)}
}
\  q^{-{\frac{3}{2}}}  (1-q)^{-{\frac{1}{2}}} \ .
\label{r2}
\end{equation}

We have obtained
 expressions (\ref{r1},\ref{r2})    by making an approximation which consists in  replacing the  problem  of the directed polymer on a tree   by a generalized random energy model (GREM) \cite{grem1,grem2,Coo}. The details of these calculations, which are  too  long to be given here, will be described  in a forthcoming paper.   In this letter we will  only present
an analytic argument   
leading to (\ref{r1},\ref{r2}) 
and numerical calculations which support these expressions. For a reason discussed below, we believe
that the  GREM approximation gives  not only the right $t$ and $q$ dependence
 in (\ref{r1},\ref{r2})
 but also the correct prefactors. 

\section{ Binary tree recursions}
{\blue Before discussing  our argument based on the GREM approximation we show  how  to determine numerically  the distribution of overlaps $P(q)$  by  recursion relations.}
For a given realization of the energies $\epsilon_b$ on the tree
the partition function satisfies the following recursion
\begin{equation}
Z_{t+1} = e^{-\beta \epsilon_1} Z_{t}^{(1)} + 
 e^{-\beta \epsilon_2} Z_{t}^{(2)} 
\label{recur}
\end{equation}
where $\epsilon_1$ and $\epsilon_2$ are distributed according to  the density $\rho(\epsilon)$ and $
Z_{t}^{(1)} $ and  
 $ Z_{t}^{(2)} $ are the partition functions of two independent copies of   a tree of height $t$. Initially $Z_0=1$.
If one introduces the generating function $G_t(x)$ defined by
$$G_t(x) = \left\langle \exp \left[-  e^{-\beta x } Z_t  \right] \right\rangle $$
where the average is over all the random energies $\epsilon_b$,
it  is easy to see using (\ref{recur}) that  it satisfies
\begin{equation}
G_{t+1} (x) = \left[\int \rho(\epsilon) d \epsilon \  G_t(x+\epsilon) \right]^2  \equiv \big[\langle G_t(x+\epsilon) \rangle_\epsilon \big]^2 
\label{Geq}
\end{equation}
 where $\langle . \rangle_\epsilon$ denotes an average over {\blue a single } random energy $\epsilon$. {\blue As $Z_0=1$} the initial condition is

\begin{equation}
G_0(x)= \exp \left[-  e^{-\beta x }   \right]  \ . 
\label{G0}
\end{equation}
{\blue One can notice \cite{spohn} that the recursion (\ref{Geq}) on $G_t$ does not depend  on the inverse temperature $\beta$. The temperature enters only in the initial condition (\ref{G0})}.

Now to calculate the distribution $P(q)$ of overlaps, it is convenient to introduce   the probability $Q_t(r)$ that two paths $i,j$  have an overlap   such that   $t  q_{i,j} \ge r$ (obviously $Q_t(0)=1$). We are going to  show that \\
\begin{equation}
\label{QH}
 Q_t(r)  ={  \frac{2^{r}}{\beta}} \int_{-\infty}^{\infty} dx  \ H_{t}^{( t-r)}(x) 
\end{equation}
where $H_{t}^{( s)}(x)$ is obtained 
by the following recursion {\blue for $t \ge s$}
\begin{equation}
H_{t+1}^{(s)}(x)=
\langle G_{ t}(x+\epsilon)\rangle_{\epsilon}  
\ \langle H_{ t}^{( s)}(x+\epsilon)\rangle_{\epsilon }
\label{Hs}
\end{equation}
{\blue with the initial condition }
\begin{equation}
H_{s}^{( s)}(x)= e^{-\beta x} 
   \ \frac{d}{dx} \left( e^{ \beta x } \frac{d}{dx}
 G_{ s}(x)\right)
\label{Hr}
\end{equation}
({\blue using the fact that $G_t(\infty)=1$ and $G_t(-\infty) =0$ one can easily check} that $Q_t(0)=1$).
This leads to   
\begin{equation}
\text{Prob}\left(q=\frac{r}{t} \right)  = Q_t(r) - Q_t(r+1) 
\label{PQr}
\end{equation}
(with  the convention that $Q_t(t+1)=0$)
{\blue allowing one to determine $P(q)$ by}
\begin{equation}
P(q) dq = P(q) \times \frac{1}{t} = 
\text{Prob}\left(q=\frac{r}{t} \right) 
\label{PQbis}
\end{equation}
{\blue since increasing $r$ by $1$ changes $dq$ by $1/t$}.

 To justify  (\ref{QH},\ref{Hs},\ref{Hr}) one can  consider the following  modified partition function
$$\Xi_{t}^{(r)} (\phi_1, \cdots \phi_{2^r}) = \sum_{j=1}^{2^r}
e^{\beta \phi_j- \beta X_j(r)} \  Z_{t-r}^{(j)} $$
where  the $X_j(r)$ are the $2^r$ energies of a tree of height $r$ and
the {\blue $Z_{t-r}^{(j)} $} are $2^r$ independent partition functions.
It is easy to see that
\begin{align}
	Q_t(r) &= \left\langle \frac{ \sum_j \big[e^{- \beta X_j(r) } \  Z_{t-r}^{(j)} \big]^2  }{
		\big[\sum_j e^{-\beta X_j(r) } \  Z_{t-r}^{(j)} \big]^2}  \right\rangle \nonumber
	\\
	&=    -  
	\frac{1}{\beta^2}
	\sum_{j=1}^{2^r} 
	\frac{\partial}{\partial \phi_j} \nonumber 
	\\
	& \left. \left(
	e^{-\beta \phi_j} 
	\  \frac{\partial}{\partial \phi_j}  \big\langle \log \Xi_{t}^{(r)}   (\phi_1, \cdots \phi_{2^r}) \big\rangle \right)  \right|_{ \phi_1 = \cdots  \phi_{2^r} = 0}
	\label{QXi}
\end{align}

This expression shows that  the calculation can be limited to the case where only one  $\phi_j$, say $\phi_1$,  is non zero. \\
 Then it is easy to check that  ${\cal H}_t^{(s)}$ defined by
\begin{equation}
{\cal H}_t^{( s)}(x,\phi_1) = 
  \left\langle \exp \left[-  e^{-\beta x }  \ \Xi_t^{( t-s)} (\phi_1,0 \cdots 0)  \right] \right\rangle 
\label{Ht-def}
\end{equation}
satisfies  for {\blue  $t \ge s$}  the following recursion
\begin{equation}
{\cal H}_{ t+1}^{( s)}(x,\phi_1) = 
  \langle G_{ t}(x+\epsilon)\rangle_{\epsilon}  
\ \langle {\cal H}_{ t}^{( s)}(x+\epsilon,\phi_1)\rangle_{\epsilon }
\label{Hsbis}
\end{equation}

with the initial condition 
\begin{equation}
 {\cal H}_{ s}^{( s)}(x,\phi_1) =  G_{ s}(x-\phi_1) \ . 
\label{Hr-initial}
\end{equation}
Then {\blue  (using the identity $\log x = \int_0^\infty (e^{-u} - e^{-ux}) du /u $) one can see } that
\begin{align}
\label{log}
& \big\langle \log \Xi_t^{(r)}   (\phi_1,0 \cdots 0) \big\rangle  =
\ \ \ \ \ \ \ \ \ \  \\
&     \ \ \ \ \ \ \ \ \ \ \int_0^\infty \frac{e^{-u} - 
\left\langle \exp\left[-u \log \Xi_t^{(r)} \right] \right\rangle}{ u} du \nonumber 
  \\ & \ \ \ \ \ \ \ \ \ \  = \beta \int  \big[G_0(x)  - {\cal H}_t^{( t-r)}(x,\phi_1) \big]  \ dx
\nonumber 
\\ & \nonumber 
\end{align}
The last step to  complete the   derivation of 
(\ref{QH},\ref{Hs},\ref{Hr}) is to take {\red  two derivatives with respect to $\phi_1$} in (\ref{Hr-initial},\ref{Hsbis},\ref{log}). 

\section{Comparison with the predictions (6,7)}
\begin{figure}
	\onefigure[width=0.4\textwidth,clip=]{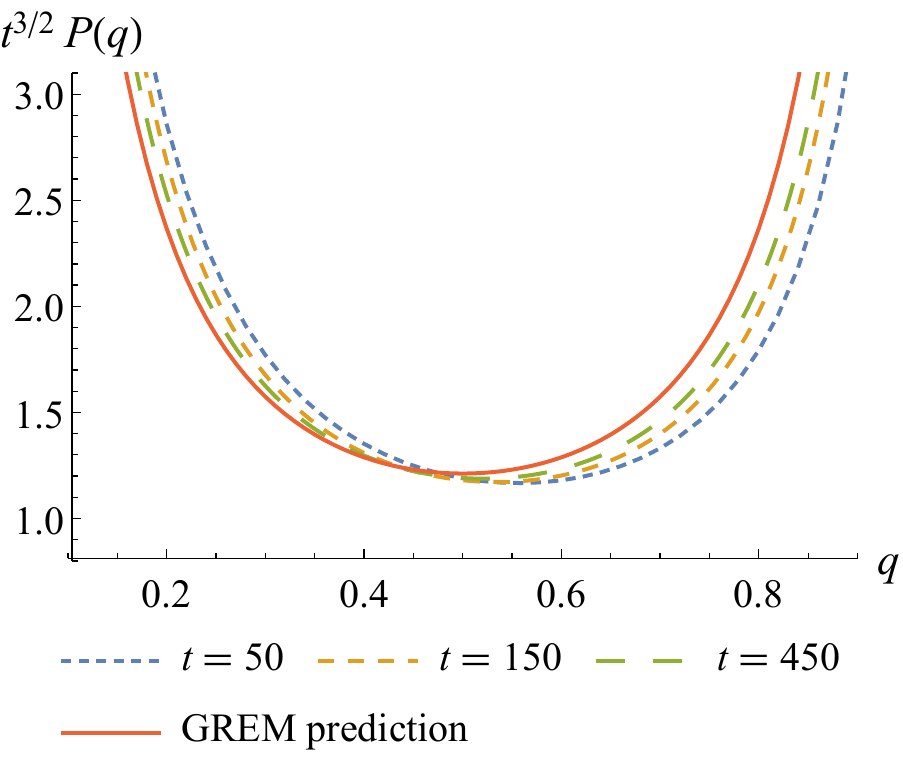}
	\caption{ \blue The  continuous line represents  the prediction (\ref{r1}) while   the dashed curves show the rescaled  distribution of overlap $t^\frac{3 }{ 2} P(q)$ determined by the recursion relations 
 (\ref{Geq},\ref{G0},\ref{QH},\ref{Hs},\ref{Hr},\ref{PQr},\ref{PQbis}) 
   at half the transition temperature ($\beta=2 \beta_c$). As $t$ increases, the agreement looks better and better.
}
	\label{fig-half-tc}
\end{figure}

\begin{figure}
	\onefigure[width=0.4\textwidth,clip=]{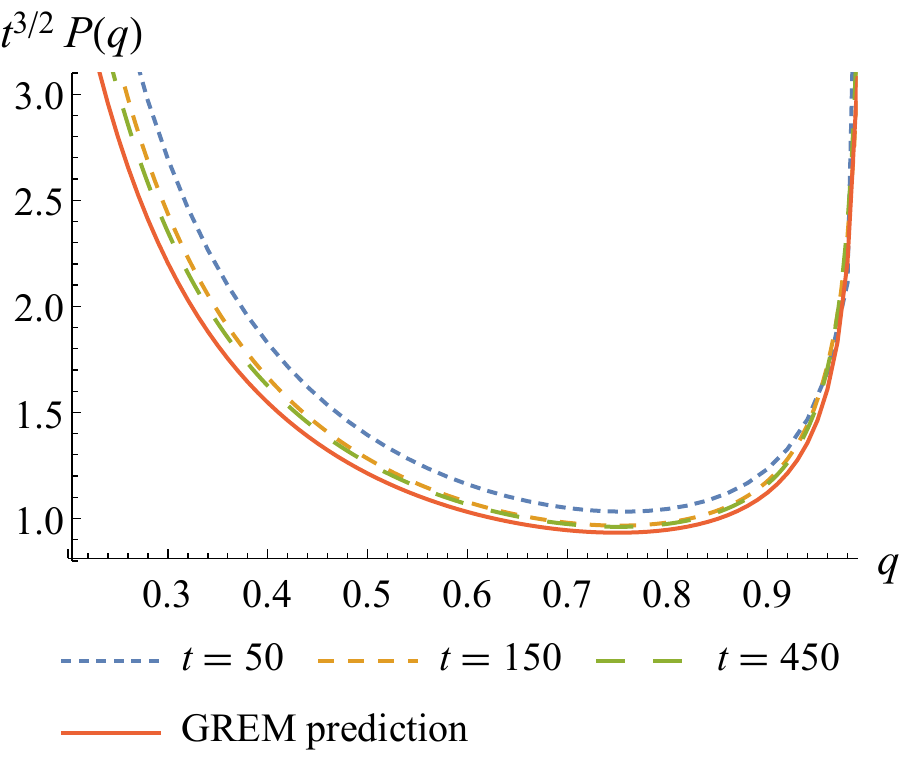}
	\caption{ \blue The same as in figure 
	\ref{fig-half-tc}
but at the transition temperature ($\beta=\beta_c$). Here again the results of  the numerical integration  of  recursions 
 (\ref{Geq},\ref{G0},\ref{QH},\ref{Hs},\ref{Hr},\ref{PQr},\ref{PQbis}) 
 seem to converge to the prediction (\ref{r2}) as $t \to \infty$.}
	\label{fig-at-tc}
\end{figure}

{\blue The dashed lines on} figures  
	\ref{fig-half-tc} and \ref{fig-at-tc} 
  show the rescaled distributions $ t^{3/2} P(q)$ versus $r/t$ obtained by 
a numerical integration of 
{\blue  (\ref{Geq},\ref{G0},\ref{QH},\ref{Hs},\ref{Hr},\ref{PQr},\ref{PQbis}) }
for  $t=50,150$ and $450$  and  {\blue a $\rho(\epsilon)$ of the form
$$\rho(\epsilon) = p \ \delta(\epsilon +1)\  +\  (1-p)\  \delta(\epsilon) \ . $$
We chose  $p=(2-\sqrt{3}) / 4 
\simeq .0669873$. 
For this choice $\beta_c = 2 \log( 2 +\sqrt{3})  \simeq 2.63392$, $v(\beta_c)=1/2$, $v'(\beta_c)=0$, $ \beta_c v''(\beta_c)=1/4 $. (The reason for chosing this value of $p$ is that $\beta_c$ has an  exact expression. For other choices of $p$  one can determine $\beta_c$   by solving   numerically 
the equation 
 $v'(\beta_c)=0$. Doing so we obtained results very similar to the ones shown in figures 
	\ref{fig-half-tc} and \ref{fig-at-tc}.)
As $t$ increases
the results of the numerical integration of the recursion relations 
 (\ref{Geq}-\ref{PQbis}) 
seem to converge to the predictions of (\ref{r1},\ref{r2}) represented by the continuous curves.}

\section{The  GREM approximation}

Let us now explain {\blue how  (\ref{r1},\ref{r2}) can be derived by an argument based on a GREM approximation}.
The approximation consists in replacing the tree of height $t$ {\blue of figure \ref{tree-fig} by a "renormalized" tree}  of height $\tau= t/n$ with
 at each level $2^n$ branches and with,  on  each bond   $b$ of this {\blue renormalized} tree, an energy $E_b$ which is the sum of $n$ independent energies $\epsilon$ distributed according to $\rho(\epsilon)$.
Varying $n$ interpolates between the original binary tree ($n=1$) and a random energy model
($n=t$) \cite{Scm}. 
(We repeated the numerical calculations shown in figures 
	\ref{fig-half-tc} and \ref{fig-at-tc} 
 in the case $n=2$  {\blue and, for large enough $t$, the  numerical} results were indistinguishable on the figure from {\red those }{\blue for } $n=1$.  {\blue As (\ref{r1},\ref{r2}) (obtained {\blue as we will see} by taking first the limit $n \to \infty$ {\red and } then the limit $t \to \infty$}) {\blue agree  so well with the numerical data of figures
	\ref{fig-half-tc} and \ref{fig-at-tc}, }
 we believe that all values of $n$, as long  as $\tau$ remains large, give the same results {\blue  (\ref{r1},\ref{r2})}  in the limit $t \to \infty$).

In the GREM approximation, $\tau$ is fixed and we take the limit of large $n$.
Then we take the large $\tau$ limit at the end.

In the low temperature phase (that is for $\beta \ge \beta_c$) and at each level of the GREM, only the energies sufficiently close to the minimal energy (at {\red  distance of order $\sqrt{n}$ when $\beta \ge \beta_c$}) contribute.  We are going to show that {\blue at  level  $p$ } from the top of  the GREM tree of height $\tau$,   these  energies,
 close to this minimal energy, are well described  by the points of
 a Poisson point process whose density is 
\begin{equation}
\widetilde{\rho}_p(E) \simeq 
\ e^{\beta_c (E-E{p})} \  F_{p}(E-E_{p}) 
\label{rhop}
\end{equation}
where 
{\blue 
\begin{align}
\label{F}
F_p(E)&=
\left(\frac{B}{\pi}\right)^\frac{p}{2}  
  \int_0 ^\infty  d x_{p-1} \cdots 
  \int_0 ^\infty  
d x_1  \nonumber \\
& \times \exp[-B \Big( (E-x_{p-1})^2 + (x_{p-1}-x_{p-2})^2 + \nonumber \\
& \cdots (x_2-x_1)^2 + x_1^2 \Big) 
\end{align}
and 
\begin{equation}
E_p= - p  n  v(\beta_c)  \ \ \ \ \ ; \ \ \ \ \ B= \Big(2 n  \beta_c v''(\beta_c)\Big)^{-1} \ .
\label{EB}
\end{equation}
}

We will need below  {\blue the following two large $p$}  asymptotics  

{\blue
\begin{equation}
F_p(0) = \sqrt{\frac{B}{\pi}} \  p^{-3/2}
\label{asympt1}
\end{equation}
}

{\blue
\begin{equation}
 \int_0^\infty F_p(E) dE 
= \frac{(2p)!}{4^p \ (p!)^2}  \simeq \frac{1}{\sqrt{\pi}} \ p^{-1 / 2}
\label{asympt2}
\end{equation}

(the  expressions  of the integrals (\ref{asympt1},\ref{asympt2}), valid for all $p \ge 1$, are exact \cite{EB}. The large $p$  asymptotics were already  known (see appendix  B of \cite{Coo})).
}

To justify (\ref{rhop},\ref{EB},\ref{F})  we note that, for large $n$, the energy $E$  on each bond of the GREM is 
the sum of $n$ random variables $\epsilon$. The  distribution  {\blue of this energy $E$} is characterized by a large deviation function $f(\epsilon)$
\begin{equation}
P(E= n \epsilon)  \ dE \sim \frac{1}{\sqrt{n \pi f''(\epsilon)}} \ e^{-n f(\epsilon)} \ dE
\label{fe}
\end{equation}
{\blue which} can be written in a parametric form  ($\epsilon= -g'(\beta) $ and $f(\epsilon) = - g(\beta) + \beta g'(\beta) $) in terms of the generating function $g(\beta)$ defined in (\ref{vbeta}) {\blue (see the appendix}). For $2^n$ independent such energies, the minimal one is close to an energy ${\blue E_1} = n \epsilon_c$ such that
\begin{equation}
\log 2 - f(\epsilon_c) =0  \ , \label{ec}
\end{equation}
 {\blue (i.e.  $\epsilon_c=-g'(\beta_c)$ where $\beta_c$ is solution of $v'(\beta_c)=0$). Near this energy, in the range $|E-{\blue E_1} |\lesssim O (\sqrt{n}) $, } one can approximate $P(E)$ by
{\red 
\begin{equation}
P(E) \simeq \sqrt{\frac{B}{\pi}} \ e^{ \beta_c (E- {\blue E_1})- B(E-{\blue E_1})^2}  \ . 
\label{P(E)}
\end{equation}
(see the appendix).}
Then one can see that expressions (\ref{rhop},\ref{EB},\ref{F}) follow simply from the recursion
$$\widetilde{\rho}_{p+1}(E) = \int_{E_{p}-A}^\infty P(E-E')\  \widetilde{\rho}_{p}(E') \ d E'  $$
{\blue with the initial condition}
$$\ \ \ \ \widetilde{\rho}_{1}(E) = P(E)  $$
where  $A$ is a cut-off $1 \ll A \ll \sqrt{n}$  \ (for example $A = n^\frac{1}{4}$)   introduced to take into account that, for large $n$,  with a probability very  close  to $1$, there is no energy $E' < E_{p}-A$. {\blue Because {\blue $B \sim n^{-1}$}  in (\ref{EB}), it is legitimate, when $n$ is large,  to  replace  this cut-off by 0 in (\ref{F}).}

\section{Derivation of  expressions (6,7) of  $P(q)$ in the  GREM approximation}

Now  the picture is that {\blue for two paths to merge   at   level}  $p$ from the top of the tree  with  $1 \le p \le (\tau-1) $, (i.e to get an overlap $p/\tau \ne 0,1$) for a GREM of height $\tau$, one need  that, due to a fluctuation,  the minimal energy at level $p$  {\blue takes}  a value of order $\log n$  lower than its typical value. {\blue A similar idea was already }  used  to understand the fluctuations of the position or the genealogy  of branching random walks with selection \cite{Bru1,Bru3} and corrections to the position of a traveling wave \cite{mu}. To be quantitative let $\psi(y)$ be the probability  of finding, at {\blue level}  $p $ from the top of the tree,   a   minimal energy $E_{p} + y$.
One has
\begin{align}
\psi(y) &= \widetilde{\rho}_{p}(E_{p}+y) \ \exp \left[ -\int_{-\infty}^y  \widetilde{\rho}_{p}(E_{p}+y') dy' \right]   \nonumber\\
&\simeq e^{\magenta{\beta_c \, y}} F_{p}(0)
\label{psi}
\end{align}
(because we anticipate that $y \sim - \log n$, one can drop the exponential and evaluate {\red (\ref{rhop})  using} the fact that $F_{p}(y) \simeq F_{p}(0)$.)

Conditioned on this fluctuation {\blue (of having a minimal energy $E_p+y$ at level $p$ from the top)} the energies at the bottom of the tree  are the points of a Poisson process whose density is
\begin{align}
 \widetilde{\rho}^\text{total} (E) &=  
 \widetilde{\rho}_{\tau- p} (E-E_p-y)  
+ \widetilde{\rho}_\tau (E) \nonumber \\
&\sim\  e^{\beta_c(E-{\blue E_\tau})}\ \left[ e^{-\beta_c y} F_{\tau-p}(E-{\blue E_\tau} + y) \right. \nonumber \\
& \qquad \left. + \ F_\tau(E-{\blue E_\tau}) \right] 
\label{2-densities}
\end{align}
where the first  term represents the subtree generated by the minimal energy $E_p+y$ at level $\tau-p$ and the second term represents the rest of the tree.

Then {\blue averaging over all realizations of the Poisson process and on all the paths gives
for  the probability that two paths  (of the subtree) chosen according to their Boltzmann weights merge at level $p$ 
\begin{align}
\label{ov}
& {\rm Pro}\left(q=\left.\frac{p}{\tau} \right| y \right) =
\int_0^\infty u \, du \int dE' \  \int dE'' 
 \\ \nonumber 
&\times \widetilde{\rho}_{\tau-p} ( E'-E_p-y) \  \widetilde{\rho}_{\tau-p} ( E''-E_p-y)\  e^{-\beta (E'+E'')} \nonumber \\
 & \times \  \exp \left[-u \left( e^{-\beta E'} + e^{-\beta E''}\right) \right. \nonumber \\
 & 
\ \ \ \ \ \ \ \ \ \ \ \ 
\left.  + \int dE \   (e^{-u \, e^{-\beta E}} -1)\  \widetilde{\rho}^\text{total} (E)  \right]  \ .
\nonumber
\end{align}
where we have used the identity $\frac{X^2}{(X+Y)^2}= \int_0^\infty u du X^2 e^{-u(X+Y)}$.}

To complete the derivation of (\ref{r1},\ref{r2})  one has to  {\blue discuss separately the two cases $\beta >  \beta_c$ and $\beta=\beta_c$}:
\\
For $\beta> \beta_c$ only energies $E$  at distances  ${\blue |E-E_\tau|  } = O(\log n) \ll \sqrt{n} $
 contribute and one can approximate 
{\red 
the expression (\ref{2-densities}) of }
 $\widetilde{\rho}^\text{total} (E) $ by
{\blue
$$ \widetilde{\rho}^\text{total} (E) 
  \sim\  e^{\beta_c(E-{\blue E_\tau})}\ \left[ e^{-\beta_c y} F_p(0) \ + \ F_\tau(0) \right] $$
where $E_\tau=- n \tau v(\beta_c) = - t v(\beta_c)$.
This leads to} (\ref{ov})  
 $${\rm Pro}\left(q=\left.\frac{p}{\tau} \right| y \right) =  \frac{\beta_c}{\beta} \left(\frac{
 F_{\tau-p}(0) e^{-\beta_c \, y}}{F_{\tau-p}(0) e^{-\beta_c \, y}  + F_\tau(0)}\right)^2 .$$
On the other hand for $\beta=\beta_c$, the integrals over energies in
 (\ref{ov}) 
 are dominated by the range $E-{\blue E_\tau}=O(\sqrt{n})$ and one gets

\begin{align*}
	&{\rm Pro}\left(q=\left.\frac{p}{\tau} \right| y \right) =  \nonumber \\
	& \frac{\beta_c}{\beta} \left(\frac{
		e^{-\beta_c \, y} \int_0^\infty F_{\tau-p}(E)dE 
	}{
		e^{-\beta_c \, y} \int_0^\infty F_{\tau-p}(E)dE 
		+ \int_0^\infty F_\tau(E) dE } \right)^2
\end{align*}

Then averaging over $y$  using the pdf  (\ref{psi})
{\blue for $\beta> \beta_c$}
 leads to
\begin{align}
	{\rm Pro}\left(q=\frac{p}{\tau} \right) & =  \frac{ {\magenta 1} }{\beta} \frac{
		F_{\tau-p}(0)  \ F_{p}(0)  }{F_\tau(0)} \nonumber \\
	&\simeq \frac{ {\magenta 1} }{\beta}
	\ \frac{1}{\sqrt{2 \pi n \beta_c v''(\beta_c)}} \ \left(\frac{\tau }{p (\tau-p)} \right)^\frac{3}{2} 
\label{A1}
\end{align}

and for $\beta=\beta_c$
  \begin{align}
  	{\rm Pro}\left(q=\frac{p }{\tau} \right) & =  \frac{{\magenta 1}  }{\beta_c} {
  		\frac{F_{\tau-p}(0)
  		\ \int_0^\infty F_p(E) dE
  	}{
  		\int_0^\infty F_\tau(E) dE  }} \nonumber \\
  	& \simeq   \frac{{\magenta 1}  }{ \beta_c}
  	\ \frac{1 }{\sqrt{2 \pi n \beta_c v''(\beta_c)}} \ \left(\frac{\tau }{p^3 (\tau-p)} \right)^\frac{1}{2}
\label{A2}
  \end{align}

where we have used the value  {\blue (\ref{EB})  of $B$ } and the asymptotics (\ref{asympt1},\ref{asympt2}).
{\blue
The last step  to obtain (\ref{r1},\ref{r2}) is to    replace $\tau$ by $\frac{t}{n}$ and  $p$ by $q \tau $ in
the above expressions  (\ref{A1},\ref{A2})   and the fact that 
$${\rm Pro}\left(q=\frac{p }{ \tau} \right) = \int_q^{q + \frac{1}{\tau}} P(q') dq' \simeq \frac{1}{\tau} P(q) \ . $$
 } 

\section{Branching Brownian motion}

{\blue 
It is straightforward to repeat the argument in the case of branching Brownian motion.
A difference is that the number ${\cal N}(t)$ of end points at the bottom of the tree  {\blue fluctuates
but if} the $X_i(t)$ are the positions of these end points  of this branching Brownian motion,  the analysis is essentially the same.  If we take a branching Brownian motion  characterized by a branching  rate $1$ (i.e. which has a probability $dt$ of branching during each infinitesimal time interval $dt$) and which diffuses according to $\langle dx^2\rangle = 2 dt$,  the only changes are that   {\red $G_t(x)$ and $H_t(x)$} become continuous functions  of time $t$ with the evolution equations  (\ref{Geq},\ref{Hs}) replaced by  

$$\frac{d G }{dt } = \frac{d^2 G }{dx^2} + G^2 -G$$

which is the Fisher-KPP equation \cite{mckean}
and

$$\frac{d H }{ dt } = \frac{d^2 H }{ dx^2} + (2G-1) H \ . $$

The velocity
$v(\beta)$  in (\ref{vbeta})  is then replaced by
$$v(\beta)=\beta + \frac{1 }{\beta}  \ .$$
With this choice, {\red one has}  $\beta_c=1, v(\beta_c)=2,  v''(\beta_c)=2$  and we expect  (\ref{r1},\ref{r2}) to hold {\red  for two end points of branching Brownian motion} sampled according to (\ref{Wi}) have their most recent common ancestor at time $qt$.

The distributions (\ref{r1},\ref{r2}) are characteristic of the large $t$ properties of the edge of the branching Brownian motion
\cite{Bru4,Bru5,Ai,Ar1,Ar2,Mallein,dean}.
If instead of choosing the end points sampled by (\ref{Wi}), we  take  any pair of points close to the leftmost particle, for example the two leftmost points of the branching Brownian motion, or the 5th and the 7th leftmost particles, expression (\ref{r1}) would hold (up to the prefactor $\beta_c/\beta$ which  would not be there).

\section{Conclusion}
In the past, the understanding of finite size  corrections and of the contribution of fluctuations in systems with   one replica symmetry breaking has been a puzzling question \cite{nie,fer,camp}.
In the present work  we have computed, without recourse to the replica trick,   the finite size corrections of the distribution of overlaps of the mean field version of the directed polymer. Our results extend  those  obtained last year \cite{peter}  on the random energy model to a more complex case.
Although  the random energy model and the directed polymer have exactly the same distribution of overlaps (\ref{P(q)}) in the thermodynamic limit, and are both representative examples of systems with one broken symmetry of replica,  their finite size corrections are quite different: here they are of order $t^{-\frac{1 }{ 2}}$ while for the random energy model they are of order $t^{-1}$.  Trying to see whether  our result (\ref{r1},\ref{r2}) could be recovered by the replica approach  is in our opinion an interesting open  question. 

{\red
Also   determining the genealogies of directed polymers in a  non-mean field case, 
in particular in  1+1 dimension, where  major progresses have been  done recently
\cite{Corw,Dot,Sasa,Joha,Ledoussal}, would be an interesting extension of the present work.
}

A more systematic way of deriving (\ref{r1},\ref{r2}), the generalisation   to calculate the overlaps of 3 or more paths and to cases where the  distribution $P(q)$, in the large $t$ limit, is {\red continuous} rather than discrete as in (\ref{P(q)}) will be  published in a forthcoming  work.}

\section{Appendix}
In this appendix we explain the relations between the large deviation function (\ref{fe}), the generating function (\ref{gbeta}) and the velocity (\ref{vbeta}).
If $E$ is the sum of $n$ random variables  $\epsilon$ distributed according to a distribution
$\rho(\epsilon)$ one has
$$\left\langle e^{-\beta E} \right\rangle = e^{n g(\beta)}$$
with $g(\beta)$ given by (\ref{gbeta}).
As 
$$\left\langle e^{-\beta E} \right\rangle = \int P(E) e^{-\beta E} dE$$
one has for large $n$
$$P(E) \simeq \sqrt{\frac{n f''\left(\frac{E }{ n} \right) }{2 \pi }} \exp \left[ -n f\left(\frac{E}{n}\right) \right]  $$
with 
\begin{equation}
g(\beta) = \max_{\epsilon} [ - \beta \epsilon - f(\epsilon)]  \ . 
\label{leg}
\end{equation}
If one knows $g(\beta)$ one can  obtain
 $f(\epsilon)$ in a parametric form
\begin{equation}
\epsilon=- g'(\beta) \ \ \ \ \ ; 
\ \ \ \ \  f(\epsilon) =- g(\beta) + \beta g'(\beta)  
\label{par}
\end{equation}
by inverting the Legendre transform (\ref{leg}).
This implies in particular that
\begin{equation}
\label{fpp}
f'(\epsilon)= -\beta \ \ \ \ ; \ \ \ \ f''(\epsilon)= \frac{1}{g''(\beta)} 
\end{equation}

In order to determine the solution $\epsilon_c$ of (\ref{ec})
$$\log 2 - f(\epsilon_c) =0  $$
one has to find the value $\beta_c$ such that
$$ \log 2  + g(\beta_c) - \beta_c \,  g'(\beta_c)=0$$
i.e. the solution of $$v'(\beta_c)=0$$
where $v(\beta)$ is defined as in (\ref{vbeta})
 $$v(\beta) = 
\frac{ 
 g(\beta) + \log 2  }{ \beta}
\ .
$$

Therefore near $\epsilon_c$ one has
\begin{align*}
f(\epsilon) &=\log 2 + f'(\epsilon_c)   (\epsilon-\epsilon_c)  + \frac{1 }{ 2} f''(\epsilon_c)   (\epsilon-\epsilon_c)^2 + \cdots 
\\
 &=\log 2 \ - \beta_c (\epsilon-\epsilon_c)  + \frac{1 }{ 2  g''(\beta_c) } (\epsilon-\epsilon_c)^2 + \cdots 
\\
& =\log 2 \ - \beta_c (\epsilon-\epsilon_c)  + \frac{1  }{2 \ \beta_c \   v''(\beta_c) } (\epsilon-\epsilon_c)^2 + \cdots 
\end{align*}
{\red and replacing $\epsilon$ by $E/n$ and $\epsilon_c$ by $E_1/n$ one gets (\ref{P(E)}) and (\ref{EB}) using the fact that $g''(\beta_c)= \beta_c v''(\beta_c)$.}

\acknowledgments
We thank Eric Brunet and Bastien Mallein for useful discussions. We would also like to thank the Higgs Centre and the Institute for Condensed Matter and Complex Systems at the University of Edinburgh for their kind hospitality.

\end{document}